\documentclass[twocolumn, aps, prb, floatfix]{revtex4}
\usepackage{graphicx}

\begin{document}
\title{Controlling a Quantum System via its Boundary Conditions}

\author{Christian Duffin$^*$}
\author{Arend G. Dijkstra$^{**}$}

\affiliation{(*) University of Leeds, School of Physics and Astronomy,
Leeds LS29JT, United Kingdom}
\affiliation{(**) University of Leeds, School of Chemistry and School of Physics and Astronomy,
Leeds LS29JT, United Kingdom, e-mail: a.g.dijkstra@leeds.ac.uk}

\begin{abstract}
We numerically study a particle in a box with moving walls. In the case where the walls are
oscillating sinusoidally with small amplitude, 
we show that states up to the fourth state can be populated with
more than 80 percent population, while higher-lying states can also be selectively 
excited. This work introduces a way of controlling
quantum systems which does not rely on (dipole) selection rules.
\end{abstract}

\maketitle

{\it Introduction} - Reliably steering a quantum system from the ground state into a 
specific state is a well-known goal in quantum technology.\cite{Haase18} 
Techniques based on laser irradiation such as Rabi oscillations or 
stimulated Raman adiabatic passage (STIRAP)\cite{Vitanov17} 
are examples that have been designed for two- or three-level systems. 
In these methods, light directly interacts with one or multiple transitions 
in the quantum system. Transitions between the ground state and the desired state
can be driven directly, or, intermediate states can be used.
In the context
of Floquet theory for periodically driven systems, 
dipolar external forces have also been considered.\cite{Bukov15, Holthaus15} 
These
techniques require either resonances with (dipole) allowed transitions or high-intensity 
external fields to be able to transfer a large amount of population.

Here, we introduce an alternative way of populating arbitrary states in a multi-level
quantum system, which operates by varying the boundary conditions in time. 
Related to this idea, in the context of optical lattices, ac modulation of the 
lattice depth,\cite{Stoeferle04, Sarkar15} 
and a harmonic trap with large-amplitude fluctuations in the 
frequency\cite{Lignier09}
have been considered. In this letter, 
in which we focus on transferring population, we look at the well-known
particle in a box system as an example. The boundary condition to be varied in this case
is the length of the box, which is changed by moving one of the walls.

The particle in a box is a well-understood quantum system, and is largely
used as a heuristic system in standard introductory textbooks.\cite{Rae05, Griffiths18}
But extending to the case of time-dependent boundaries, the literature is
largely mathematically driven.
For any motion that is slow, the adiabatic approximation will suffice,\cite{Pinder90}
in which the expansion coefficients ($C_n$ defined later) of the system can be assumed to be
time-independent. Because this assumption does not hold generally, exact
solutions are limited to select cases.
Analytical solutions of the time-dependent Schr\"odinger equation are
known for constant wall velocity\cite{Doescher69} and for certain cases of the
quantum harmonic oscillator, in which the angular frequency of the potential
is parametrised as a specific function of the well width, $L$.\cite{Makowski91, Makowski92, MakowskiPeplowski92}
Cooney has also proposed a means of deriving solutions for accelerating
walls using extended transformation methods,\cite{Cooney17} although
these only hold in the limit that the acceleration is slow.
Employing a numerical approach,\cite{Tarr05}  we are able to study arbitrary
wall motion.

{\it Model} - The usual particle in a one-dimensional 
box, as it is introduced in standard textbooks, 
has infinitely high walls on both sides
of a box with length $L$, in which a particle with mass $m$ resides. 
Elementary solutions to the time-independent Schr\"odinger
equation give the energy levels (eigenstates) for each integer quantum number $n$ as 
$E_n = \hbar^2 \pi^2 n^2 / 2 m L^2$.
The corresponding eigenfunctions are $u_n = \sqrt{2/L} \sin n \pi x/L$,
where $0 < x < L$ is the coordinate. Any wave function $\psi$ can be expanded on the 
complete basis of
these eigenfunctions, and evolves in time according to the time-dependent Schr\"odinger
equation as
\begin{equation}
  \psi(x,t) = \sum_n C_n u_n \exp(-i E_n t / \hbar), 
\end{equation}  
where $C_n$ are the
expansion coefficients determined from the initial condition. 

The numerical solution for a system with a moving wall, described with a time-dependent
box length $L(t)$, is obtained by allowing the expansion coefficients $C_n$ to be
time-dependent.\cite{Tarr05} The details of the derivation are presented in Ref.\onlinecite{Tarr05}.
Briefly, the ansatz wave function with time-dependent expansion coefficients is plugged
into the time-dependent Schr\"odinger equation. Simplifying using the orthogonality
of the eigenstates as well as their explicit form, one arrives at the
final result as a set of coupled linear differential equations for the expansion 
coefficients
\begin{eqnarray}
  \dot C_k(t) &=& \sum_n \frac{2 (-1)^{k+n} k n}{(n^2-k^2)} \frac{\dot L(t)}{L(t)} C_n(t) \\
       &\times&      \exp \left( \frac{-i L^2(t)}{\hbar} (E_n(t) - E_k(t)) \int_0^t \mathrm{d}\tau
                         \frac{1}{L^2(\tau)}  \nonumber
                   \right).
\end{eqnarray}
In this equation, dots denote time derivatives, and the energies acquire a time dependence
through the varying box length. The set of coupled equations is solved
numerically using the Dormand-Prince algorithm as implemented in Matlab.\cite{ode45}

As the initial condition of our simulations, we will assume that the system is prepared
in the ground state. This choice could be easily generalized to superpositions of
eigenstates, which would allow,
for example, the study of quantum carpets.\cite{Berry96, Berry01, Marzoli98}

In all our calculations, we will be working in natural units, taking the mass of the
particle to be $m=1$, and $\hbar=1$.
As a consequence, our time domain will be in units of $L_0^2 m/\hbar$, where $L_0$ is
the initial or average length of the box. For a proton in
a 1 nm box, this time unit would correspond to 16 ns.

{\it Constant velocity} - We validated our numerical approach by comparing time-dependent
populations of states with those obtained form known analytical 
solutions\cite{Doescher69, MakowskiPeplowski92, Greenberger88} for a wall moving 
with a constant velocity $v$, that is,
$L(t) = L_0 + v t$. We
briefly mention the results for a contracting box (negative $v$). 
When the box length becomes
small, all state populations tend to a constant and the amplitude of oscillations
decreases. The amount of population transferred from the ground state to other states
increases with increasing speed of the wall motion. If the motion of the wall is
sufficiently slow, little population is transferred, i.e. the system behaves
adiabatically. For high enough speeds of the wall, it is possible to transfer most
of the population out of the ground state, and higher lying states can be populated
more than lower lying states. 
In this process, the particle gains momentum. 
The increase of the particle's momentum without a force
acting on it can be explained by the 'Greenberger 
effect'\cite{MakowskiPeplowski92, Greenberger88} and is a result of the delocalized
nature of the wave function. Essentially, this is the spreading out of the wave packet
as it would for a free particle. 

The states in a box with moving walls do not cross, nor do they exhibit avoided crossings. 
For a slowly varying system the adiabatic approximation is expected to be valid,
and the particle mostly stays in the same time-dependent eigenstate. 
But, a uniformly expanding box is never eternally adiabatic. However slow the expansion, 
eventually the states will be so close together in energy that non-adiabatic effects 
become important. For a uniformly contracting box, the inverse principle is also
true, which is to say that however fast the contraction, the states will eventually
be sufficiently far from one another for the system to behave adiabatically.

\begin{figure}[h]
 \includegraphics[width=8.5cm]{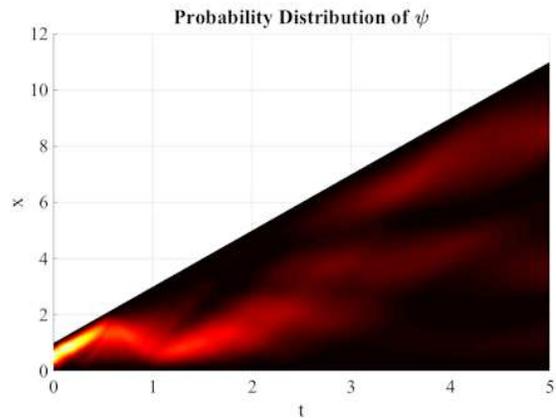}
\caption{\label{fig:6d} 
Probability distribution for a uniformly expanding well with speed $v=2$ and $L_0 = 1$.
The probability increases from black to red to yellow.
}
\end{figure}

There are several ways to visualize the particle in a box dynamics, which include
plotting the eigenstate populations $|C_n|^2$, 
the expectations values of position $\langle \psi | x | \psi \rangle$, momentum
$\langle \psi | p | \psi \rangle$,
and kinetic energy $\langle \psi | p^2/2m | \psi \rangle$, or the probability 
distribution $|\psi(x,t)|^2$.
The probability distribution for a particle in a linearly expanding
box is shown in figure \ref{fig:6d} for a speed of $v=2$. 
From the increasing number of dark lines in
this picture, it is clear that higher-lying states are populated. Such effects are even
more striking when the expansion speed is increased and are also reflected in expectation
values of position and momentum.

\begin{figure}[h]
 \includegraphics[width=8.5cm]{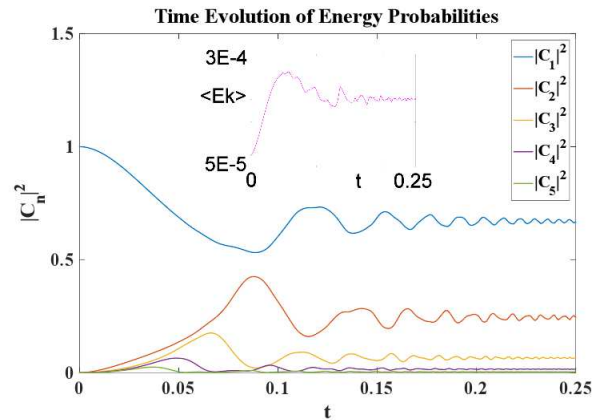}
\caption{\label{fig:8a} 
State populations for an exponentially contracting well with $v=-9$ and $L_0 = 1$.
The inset shows the kinetic energy.
}
\end{figure}

{\it Acceleration} - While uniform motion has allowed us to investigate the limits of
adiabaticity, an exponentially moving wall may be a better model for contraction. We
will use as the equation for the wall length $L(t) = L_0 \exp v t / L_0$. For
appropriately chosen (negative) $v$, 
we can initialise the system into a certain state using
non-adiabatic transitions and then force it into an adiabatic regime. In this way,
it becomes possible to prepare certain superpositions in a stable way. Results for
the population dynamics are shown in figure \ref{fig:8a}. This process could be used
to accelerate a particle.

{\it Oscillating wall} -  
Our main results are for a sinusoidally moving wall. We set the
length of the box as
\begin{equation}
  L(t) = L_0 + v \sin \omega t,
\end{equation}
with angular frequency $\omega$ and amplitude $v$, as well as average box size $L_0$.

\begin{figure*}[t]
 \includegraphics[width=17cm, height=10cm]{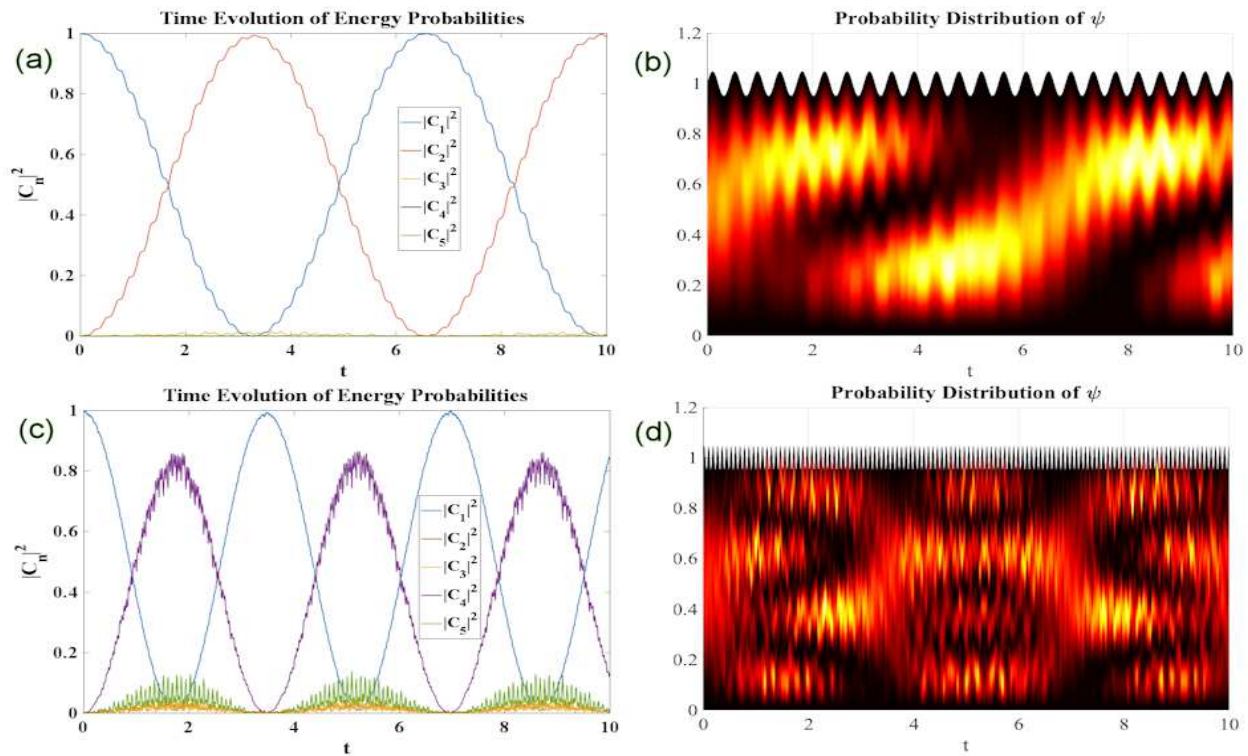}
\caption{\label{fig:14c-f} 
(a) and (c) State populations and (b) and (d) probability distributions 
for a sinusoidally moving wall with
frequency (a) and (b) 14.7605 and (c) and (d) 73.7048.
and small amplitude $v/L_0 = 0.05$ and $L_0 = 1$. The frequency in the top (bottom) row
is chosen to maximize the population in the second (fourth) eigenstate.
}
\end{figure*}

In figure \ref{fig:14c-f} we show the time dependence of the second and fourth state for a box
driven with a frequency chosen in such a way that these states acquire the maximal
possible population for the chosen driving amplitude. The figure shows that the excitation
is selectively populating the desired state, with only small populations of other
states. By doing this the system absorbs energy by increasing the kinetic energy of the 
particle. To prevent the population from going back to the ground state, as is the
case for periodic driving, one could imagine more complex driving patterns to stabilize
a desired state, for example,
following periodic motion with exponential wall motion. 

\begin{figure}[h]
 \includegraphics[width=8.5cm]{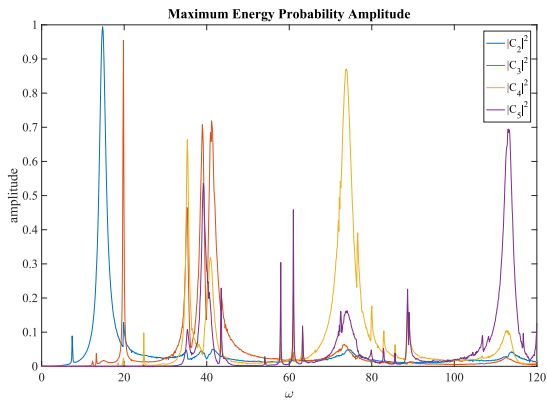}
\caption{\label{fig:13} 
  Maximum populations of each state as a function of driving frequency for a sinusoidally
  moving wall with angular frequency $\omega$, $L_0 = 1$ and $v/L_0 = 0.05$. 
  The sharp resonances allow selectively
  populating certain individual eigenstates.
}
\end{figure}

To investigate which driving frequency must be chosen to excite each state, we plot
figure \ref{fig:13}. This figure shows the highest population of the first five particle
in a box eigenstates across a time domain of $t=0$ to $t=10$ as a function of the driving frequency $\omega$.
In this figure, we observe many sharp resonances, which indicate optimal driving
frequencies. Remarkably, there are peaks for each of the states, showing that
each state can be driven to large population. The first four states all have maximum
populations above 80 percent, confirming that they can be selectively excited. Although
not shown here, we have also plotted a similar figure for even higher-lying states, and
we have confirmed that they can also be populated significantly through similar
resonances (population larger than 0.3 for all states up to $n=10$). 
We find that the resonance frequency needed to populate the second eigenstate
is $\omega=14.7605$, close to the expected value of $E_2 - E_1 = 3 \pi^2/2 = 14.8044$.
We attribute the small difference to the non-infinitesimal amplitude of the wall motion.
Resonances to higher-lying peaks cannot be explained with this simple argument. 
For completeness, we have also explored other values of the amplitude $v$ such as $v=0.04$
and $v=0.06$. As expected, we find similar resonances as in the case of $v=0.05$, but
the maximum populations can be tuned by changing $v$.

At this point, we note that populating higher-lying particle in a box states with a 
dipolar force requires high intensities.\cite{Holthaus15} 
In contrast, our method produces large populations with small amplitude driving of
the wall. Indeed, we can populate the lowest 6 states with more than 60 percent population,
and the lowest four states with more than 80 percent. 
This could lead to more efficient ways of populating such states. It is,
of course, not clear how to move an infinitely high wall, and our mechanism
should be investigated in other potentials as well.

To complete our investigation of sinusoidal motion, we have also considered motion with
low frequency\cite{Cooney17} but large amplitude. In this case, we find a similar 
interplay between
adiabatic evolution and non-adiabatic mixing as expected from the exponentially
moving wall.

{\it Experimental systems} - State selective excitation techniques such as STIRAP
have many applications.\cite{Vitanov17}
Implementation of our proposal could be attempted in optical lattices, optomechanical
resonators\cite{Mann18, Lau18} or in nano-electro-mechanical systems.\cite{Sohn18}
Furthermore, acceleration by a moving wall could be used to inject kinetic energy into 
particles.
Recently, the particle in the box has also been used as a model system
for excess protons in water,\cite{Dahms17} which is known to be a highly dynamic 
system,\cite{Thamer15} and could therefore be modelled with our approach. 

{\it Conclusion} - In conclusion, we have been able to use numerical techniques in 
order to investigate the nature of the particle in a box 
with moving walls in terms of the adiabatic and 
the non-adiabatic regimes. Through this we have demonstrated a mechanism by which the 
particle can be prepared in a stable state of tuneable energy, controlled by the speed 
of contraction. We have also described a method to selectively populate quantum states 
by driving the system's boundaries periodically with small amplitude 
by tuning the frequency.
In future work, it will be important to investigate decay processes that lead to
losses from the desired state. Techniques such as STIRAP employ coherences with an 
intermediate state without significantly populating it, therefore making the technique
insensitive to losses from this state. However, in our technique, no intermediate
state is necessary because the model does not rely on dipole allowed transitions.
Also, the particle in an infinitely deep potential well is an artificial model, and
more realistic models potentials such as a finite well should be investigated.
Extending this analysis to a stochastic regime of motion, such as Brownian 
motion,\cite{Hida80} would be useful in describing the dynamics of excess proton in water 
systems. Finally, preparing the system in superposition states to simulate quantum 
carpets,\cite{Berry01, Berry96, Marzoli98}  
it could be worth investigating how these patterns respond to the types of motion 
demonstrated in this work.

{\it Acknowledgements} - We thank Dr Zlatko Papi\'c, Prof Jiannis Pachos and Dr Marcelo
Miranda for helpful discussions.


\begin{thebibliography}{27}
\expandafter\ifx\csname natexlab\endcsname\relax\def\natexlab#1{#1}\fi
\expandafter\ifx\csname bibnamefont\endcsname\relax
  \def\bibnamefont#1{#1}\fi
\expandafter\ifx\csname bibfnamefont\endcsname\relax
  \def\bibfnamefont#1{#1}\fi
\expandafter\ifx\csname citenamefont\endcsname\relax
  \def\citenamefont#1{#1}\fi
\expandafter\ifx\csname url\endcsname\relax
  \def\url#1{\texttt{#1}}\fi
\expandafter\ifx\csname urlprefix\endcsname\relax\def\urlprefix{URL }\fi
\providecommand{\bibinfo}[2]{#2}
\providecommand{\eprint}[2][]{\url{#2}}

\bibitem[{\citenamefont{Haase et~al.}(2018)\citenamefont{Haase, Wang, Casanova,
  and Plenio}}]{Haase18}
\bibinfo{author}{\bibfnamefont{J.~F.} \bibnamefont{Haase}},
  \bibinfo{author}{\bibfnamefont{Z.-Y.} \bibnamefont{Wang}},
  \bibinfo{author}{\bibfnamefont{J.}~\bibnamefont{Casanova}}, \bibnamefont{and}
  \bibinfo{author}{\bibfnamefont{M.~B.} \bibnamefont{Plenio}},
  \bibinfo{journal}{Phys. Rev. Lett.} \textbf{\bibinfo{volume}{121}},
  \bibinfo{pages}{050402} (\bibinfo{year}{2018}).

\bibitem[{\citenamefont{Vitanov et~al.}(2017)\citenamefont{Vitanov, Rangelov,
  Shore, and Bergmann}}]{Vitanov17}
\bibinfo{author}{\bibfnamefont{N.~V.} \bibnamefont{Vitanov}},
  \bibinfo{author}{\bibfnamefont{A.~A.} \bibnamefont{Rangelov}},
  \bibinfo{author}{\bibfnamefont{B.~W.} \bibnamefont{Shore}}, \bibnamefont{and}
  \bibinfo{author}{\bibfnamefont{K.}~\bibnamefont{Bergmann}},
  \bibinfo{journal}{Rev. Mod. Phys.} \textbf{\bibinfo{volume}{89}},
  \bibinfo{pages}{015006} (\bibinfo{year}{2017}).

\bibitem[{\citenamefont{Bukov et~al.}(2015)\citenamefont{Bukov, Gopalakrishnan,
  Knap, and Demler}}]{Bukov15}
\bibinfo{author}{\bibfnamefont{M.}~\bibnamefont{Bukov}},
  \bibinfo{author}{\bibfnamefont{S.}~\bibnamefont{Gopalakrishnan}},
  \bibinfo{author}{\bibfnamefont{M.}~\bibnamefont{Knap}}, \bibnamefont{and}
  \bibinfo{author}{\bibfnamefont{E.}~\bibnamefont{Demler}},
  \bibinfo{journal}{Phys. Rev. Lett.} \textbf{\bibinfo{volume}{115}},
  \bibinfo{pages}{205301} (\bibinfo{year}{2015}).

\bibitem[{\citenamefont{Holthaus}(2015)}]{Holthaus15}
\bibinfo{author}{\bibfnamefont{M.}~\bibnamefont{Holthaus}},
  \bibinfo{journal}{J. Phys. B: At. Mol. Opt. Phys.}
  \textbf{\bibinfo{volume}{49}}, \bibinfo{pages}{013001}
  (\bibinfo{year}{2015}).

\bibitem[{\citenamefont{St\"oferle et~al.}(2004)\citenamefont{St\"oferle,
  Moritz, Schori, K\"ohl, and Esslinger}}]{Stoeferle04}
\bibinfo{author}{\bibfnamefont{T.}~\bibnamefont{St\"oferle}},
  \bibinfo{author}{\bibfnamefont{H.}~\bibnamefont{Moritz}},
  \bibinfo{author}{\bibfnamefont{C.}~\bibnamefont{Schori}},
  \bibinfo{author}{\bibfnamefont{M.}~\bibnamefont{K\"ohl}}, \bibnamefont{and}
  \bibinfo{author}{\bibfnamefont{T.}~\bibnamefont{Esslinger}},
  \bibinfo{journal}{Phys. Rev. Lett.} \textbf{\bibinfo{volume}{92}},
  \bibinfo{pages}{130403} (\bibinfo{year}{2004}).

\bibitem[{\citenamefont{Sarkar et~al.}(2015)\citenamefont{Sarkar, Sensarma, and
  Sengupta}}]{Sarkar15}
\bibinfo{author}{\bibfnamefont{S.~D.} \bibnamefont{Sarkar}},
  \bibinfo{author}{\bibfnamefont{R.}~\bibnamefont{Sensarma}}, \bibnamefont{and}
  \bibinfo{author}{\bibfnamefont{K.}~\bibnamefont{Sengupta}},
  \bibinfo{journal}{Phys. Rev. B} \textbf{\bibinfo{volume}{92}},
  \bibinfo{pages}{174529} (\bibinfo{year}{2015}).

\bibitem[{\citenamefont{Lignier et~al.}(2009)\citenamefont{Lignier, Zenesini,
  Ciampini, Morsch, Arimondo, Montangero, Pupillo, and Fazio}}]{Lignier09}
\bibinfo{author}{\bibfnamefont{H.}~\bibnamefont{Lignier}},
  \bibinfo{author}{\bibfnamefont{A.}~\bibnamefont{Zenesini}},
  \bibinfo{author}{\bibfnamefont{D.}~\bibnamefont{Ciampini}},
  \bibinfo{author}{\bibfnamefont{O.}~\bibnamefont{Morsch}},
  \bibinfo{author}{\bibfnamefont{E.}~\bibnamefont{Arimondo}},
  \bibinfo{author}{\bibfnamefont{S.}~\bibnamefont{Montangero}},
  \bibinfo{author}{\bibfnamefont{G.}~\bibnamefont{Pupillo}}, \bibnamefont{and}
  \bibinfo{author}{\bibfnamefont{R.}~\bibnamefont{Fazio}},
  \bibinfo{journal}{Phys. Rev. A} \textbf{\bibinfo{volume}{79}},
  \bibinfo{pages}{041601(R)} (\bibinfo{year}{2009}).

\bibitem[{\citenamefont{Rae}(2005)}]{Rae05}
\bibinfo{author}{\bibfnamefont{A.}~\bibnamefont{Rae}},
  \emph{\bibinfo{title}{Quantum Physics: a Beginner's Guide}}
  (\bibinfo{publisher}{Oneworld Publications}, \bibinfo{year}{2005}).

\bibitem[{\citenamefont{Griffiths}(2018)}]{Griffiths18}
\bibinfo{author}{\bibfnamefont{D.}~\bibnamefont{Griffiths}},
  \emph{\bibinfo{title}{Introduction to Quantum Mechanics}}
  (\bibinfo{publisher}{Cambridge University Press}, \bibinfo{year}{2018}).

\bibitem[{\citenamefont{Pinder}(1990)}]{Pinder90}
\bibinfo{author}{\bibfnamefont{D.}~\bibnamefont{Pinder}},
  \bibinfo{journal}{American Journal of Physics} \textbf{\bibinfo{volume}{58}},
  \bibinfo{pages}{54} (\bibinfo{year}{1990}).

\bibitem[{\citenamefont{Doescher and Rice}(1969)}]{Doescher69}
\bibinfo{author}{\bibfnamefont{S.}~\bibnamefont{Doescher}} \bibnamefont{and}
  \bibinfo{author}{\bibfnamefont{M.}~\bibnamefont{Rice}},
  \bibinfo{journal}{American Journal of Physics} \textbf{\bibinfo{volume}{37}},
  \bibinfo{pages}{1246} (\bibinfo{year}{1969}).

\bibitem[{\citenamefont{Makowski and Dembinski}(1991)}]{Makowski91}
\bibinfo{author}{\bibfnamefont{A.}~\bibnamefont{Makowski}} \bibnamefont{and}
  \bibinfo{author}{\bibfnamefont{S.}~\bibnamefont{Dembinski}},
  \bibinfo{journal}{Physics Letters A} \textbf{\bibinfo{volume}{154}},
  \bibinfo{pages}{217} (\bibinfo{year}{1991}).

\bibitem[{\citenamefont{Makowski}(1992)}]{Makowski92}
\bibinfo{author}{\bibfnamefont{A.}~\bibnamefont{Makowski}},
  \bibinfo{journal}{Journal of Physics A} \textbf{\bibinfo{volume}{25}},
  \bibinfo{pages}{3419} (\bibinfo{year}{1992}).

\bibitem[{\citenamefont{Makowski and Peplowski}(1992)}]{MakowskiPeplowski92}
\bibinfo{author}{\bibfnamefont{A.}~\bibnamefont{Makowski}} \bibnamefont{and}
  \bibinfo{author}{\bibfnamefont{P.}~\bibnamefont{Peplowski}},
  \bibinfo{journal}{Physics Letters A} \textbf{\bibinfo{volume}{163}},
  \bibinfo{pages}{143} (\bibinfo{year}{1992}).

\bibitem[{\citenamefont{Cooney}(2017)}]{Cooney17}
\bibinfo{author}{\bibfnamefont{K.}~\bibnamefont{Cooney}},
  \bibinfo{journal}{arXiv preprint} \textbf{\bibinfo{volume}{arXiv:1703.05282}}
  (\bibinfo{year}{2017}).

\bibitem[{\citenamefont{Tarr}(2005)}]{Tarr05}
\bibinfo{author}{\bibfnamefont{L.}~\bibnamefont{Tarr}}, \bibinfo{journal}{PhD
  thesis, Reed College}  (\bibinfo{year}{2005}).

\bibitem[{ode()}]{ode45}
\urlprefix\url{https://www.mathworks.com/help/matlab/ref/ode45.html}.

\bibitem[{\citenamefont{Berry}(1996)}]{Berry96}
\bibinfo{author}{\bibfnamefont{M.~V.} \bibnamefont{Berry}},
  \bibinfo{journal}{Journal of Physics A} \textbf{\bibinfo{volume}{29}},
  \bibinfo{pages}{6617} (\bibinfo{year}{1996}).

\bibitem[{\citenamefont{Berry}(2001)}]{Berry01}
\bibinfo{author}{\bibfnamefont{M.~V.} \bibnamefont{Berry}},
  \bibinfo{journal}{Physics World} \textbf{\bibinfo{volume}{14}},
  \bibinfo{pages}{39} (\bibinfo{year}{2001}).

\bibitem[{\citenamefont{Marzoli et~al.}(1998)\citenamefont{Marzoli, Saif,
  Bialynicki-Birula, Friesch, Kaplan, and Schleich}}]{Marzoli98}
\bibinfo{author}{\bibfnamefont{I.}~\bibnamefont{Marzoli}},
  \bibinfo{author}{\bibfnamefont{F.}~\bibnamefont{Saif}},
  \bibinfo{author}{\bibfnamefont{I.}~\bibnamefont{Bialynicki-Birula}},
  \bibinfo{author}{\bibfnamefont{O.}~\bibnamefont{Friesch}},
  \bibinfo{author}{\bibfnamefont{A.}~\bibnamefont{Kaplan}}, \bibnamefont{and}
  \bibinfo{author}{\bibfnamefont{W.}~\bibnamefont{Schleich}},
  \bibinfo{journal}{Acta Phys. Slov.} \textbf{\bibinfo{volume}{48}},
  \bibinfo{pages}{323} (\bibinfo{year}{1998}).

\bibitem[{\citenamefont{Greenberger}(1988)}]{Greenberger88}
\bibinfo{author}{\bibfnamefont{D.}~\bibnamefont{Greenberger}},
  \bibinfo{journal}{Physica B+C} \textbf{\bibinfo{volume}{151}},
  \bibinfo{pages}{374} (\bibinfo{year}{1988}).

\bibitem[{\citenamefont{Mann et~al.}(2018)\citenamefont{Mann, Bakhtiari,
  Pelster, and Thorwart}}]{Mann18}
\bibinfo{author}{\bibfnamefont{N.}~\bibnamefont{Mann}},
  \bibinfo{author}{\bibfnamefont{M.~R.} \bibnamefont{Bakhtiari}},
  \bibinfo{author}{\bibfnamefont{A.}~\bibnamefont{Pelster}}, \bibnamefont{and}
  \bibinfo{author}{\bibfnamefont{M.}~\bibnamefont{Thorwart}},
  \bibinfo{journal}{Phys. Rev. Lett.} \textbf{\bibinfo{volume}{120}},
  \bibinfo{pages}{063605} (\bibinfo{year}{2018}).

\bibitem[{\citenamefont{Lau et~al.}(2018)\citenamefont{Lau, Eisfeld, and
  Rost}}]{Lau18}
\bibinfo{author}{\bibfnamefont{H.-K.} \bibnamefont{Lau}},
  \bibinfo{author}{\bibfnamefont{A.}~\bibnamefont{Eisfeld}}, \bibnamefont{and}
  \bibinfo{author}{\bibfnamefont{J.-M.} \bibnamefont{Rost}},
  \bibinfo{journal}{arXiv preprint} \textbf{\bibinfo{volume}{arXiv:1803.00150}}
  (\bibinfo{year}{2018}).

\bibitem[{\citenamefont{Sohn et~al.}(2018)\citenamefont{Sohn, Meesala,
  Pingault, Atikian, Holzgrafe, G{\"u}ndo{\u g}an, Stavrakas, Stanley,
  Sipahigil, Choi et~al.}}]{Sohn18}
\bibinfo{author}{\bibfnamefont{Y.-I.} \bibnamefont{Sohn}},
  \bibinfo{author}{\bibfnamefont{S.}~\bibnamefont{Meesala}},
  \bibinfo{author}{\bibfnamefont{B.}~\bibnamefont{Pingault}},
  \bibinfo{author}{\bibfnamefont{H.~A.} \bibnamefont{Atikian}},
  \bibinfo{author}{\bibfnamefont{J.}~\bibnamefont{Holzgrafe}},
  \bibinfo{author}{\bibfnamefont{M.}~\bibnamefont{G{\"u}ndo{\u g}an}},
  \bibinfo{author}{\bibfnamefont{C.}~\bibnamefont{Stavrakas}},
  \bibinfo{author}{\bibfnamefont{M.~J.} \bibnamefont{Stanley}},
  \bibinfo{author}{\bibfnamefont{A.}~\bibnamefont{Sipahigil}},
  \bibinfo{author}{\bibfnamefont{J.}~\bibnamefont{Choi}}, \bibnamefont{et~al.},
  \bibinfo{journal}{Nature Communications} \textbf{\bibinfo{volume}{9}},
  \bibinfo{pages}{2012} (\bibinfo{year}{2018}),
  \urlprefix\url{https://doi.org/10.1038/s41467-018-04340-3}.

\bibitem[{\citenamefont{Dahms et~al.}(2017)\citenamefont{Dahms, Fingerhut,
  Nibbering, Pines, and Elsaesser}}]{Dahms17}
\bibinfo{author}{\bibfnamefont{F.}~\bibnamefont{Dahms}},
  \bibinfo{author}{\bibfnamefont{B.~P.} \bibnamefont{Fingerhut}},
  \bibinfo{author}{\bibfnamefont{E.~T.} \bibnamefont{Nibbering}},
  \bibinfo{author}{\bibfnamefont{E.}~\bibnamefont{Pines}}, \bibnamefont{and}
  \bibinfo{author}{\bibfnamefont{T.}~\bibnamefont{Elsaesser}},
  \bibinfo{journal}{Science} \textbf{\bibinfo{volume}{357}},
  \bibinfo{pages}{491} (\bibinfo{year}{2017}).

\bibitem[{\citenamefont{Th{\"a}mer et~al.}(2015)\citenamefont{Th{\"a}mer,
  De~Marco, Ramasesha, Mandal, and Tokmakoff}}]{Thamer15}
\bibinfo{author}{\bibfnamefont{M.}~\bibnamefont{Th{\"a}mer}},
  \bibinfo{author}{\bibfnamefont{L.}~\bibnamefont{De~Marco}},
  \bibinfo{author}{\bibfnamefont{K.}~\bibnamefont{Ramasesha}},
  \bibinfo{author}{\bibfnamefont{A.}~\bibnamefont{Mandal}}, \bibnamefont{and}
  \bibinfo{author}{\bibfnamefont{A.}~\bibnamefont{Tokmakoff}},
  \bibinfo{journal}{Science} \textbf{\bibinfo{volume}{350}},
  \bibinfo{pages}{78} (\bibinfo{year}{2015}).

\bibitem[{\citenamefont{Hida}(1980)}]{Hida80}
\bibinfo{author}{\bibfnamefont{T.}~\bibnamefont{Hida}},
  \emph{\bibinfo{title}{Brownian Motion}}
  (\bibinfo{publisher}{Springer-Verlag}, \bibinfo{year}{1980}).

\end{thebibliography}
\end{document}